\documentclass{article}
\usepackage{fullpage}
\usepackage{booktabs} 

\usepackage{authblk}

\usepackage{amsmath,amssymb,amsfonts,amssymb}
\usepackage{graphicx}
\usepackage{textcomp}
\usepackage{xcolor}
\usepackage{url}
\usepackage{algorithm}
\usepackage[noend]{algpseudocode}
\usepackage{float}
\usepackage{soul}

\newtheorem{definition}{Definition}

\title{A novel algorithm for clearing financial obligations between companies - an application within the Romanian Ministry of Economy}

\author[1]{Lucian-Ionut Gavrila}

\author[2]{Alexandru Popa}

\affil[1]{Faculty of Mathematics and Computer Science
University of Bucharest, Romania}
\affil[2]{National Institute for Research and Development in Informatics}
\affil[ ]{\texttt{lucian.ionut.gavrila@drd.unibuc.ro, alexandru.popa@fmi.unibuc.ro}}

\begin{document}

\maketitle

\begin{abstract}
The concept of clearing or netting, as defined in the glossaries of European Central Bank, has a great impact on the economy of a country influencing the exchanges and the interactions between companies. On short, netting refers to an alternative to the usual way in which the companies make the payments to each other: it is an agreement in which each party sets off amounts it owes against amounts owed to it. Based on the amounts two or more parties owe between them, the payment is substituted by a direct settlement. In this paper we introduce a set of graph algorithms which provide optimal netting solutions for the scale of a country economy. The set of algorithms computes results in an efficient time and is tested on invoice data provided by the Romanian Ministry of Economy. Our results show that classical graph algorithms are still capable of solving very important modern problems.
\end{abstract}

\section{Introduction}

According to the European Central Bank Glossary of terms related to payment, clearing and settlement systems~\cite{ECBglossary} the concept of netting (also known as clearing) between entities is defined as ``an arrangement among three or more parties for the netting of obligations and the settling of multilateral net settlement positions''. Companies acting in an economy express their interaction by issuing invoices one to another. These invoices imply an obligation of payment, which is normally settled through a bank transfer. Netting offers an alternative to direct payments. For example, considering two companies, a national transportation provider and a national energy provider, they interact and issue invoices to each other. Therefore, they have mutual debts which can be settled through netting, instead of a direct payment. 

In this paper, we consider a particular netting approach that involves computing circuits of debts between companies and settling them. For example, company A has to pay an amount of 32.000 USD to company B, which has to pay 23.000 USD to company C, which has to pay 25.000 USD to company A. Through a clearing system, a netting circuit is determined between companies A, B and C. The option of clearing the minimum amount of 23.000 USD is presented to them. If the companies accept the clearing, the minimum amount will be subtracted from the payment obligations. Therefore, company A remains to pay 9.000 USD to B, B has no more payment obligations to C, C remains to pay 2.000 USD to A. The remaining amounts can be considered in other netting circuits. Netting brings important advantages like reducing the delays implied by bank transfers, reducing the number of transfer transactions, providing lower costs by avoiding the transfer fees, improving the credit risk~\cite{SYSTEMIC_RISK_IN_INTERNATIONAL_SETTLEMENTS}. According to~\cite{SystemicRiskInTheDanishInterbankNettingSystem}, ``by reducing the number and overall value of payments between financial institutions, netting can enhance the efficiency of payment systems''. In financial crisis scenarios, small, medium  and in some cases even large scale companies do not have the necessary liquidity to face their financial obligations. This idea is outlined in~\cite{ConnectivityNettingAndSystemicRiskOfPaymentSystems} in a similar manner for the banking system. Chen and Wu~\cite{ConnectivityNettingAndSystemicRiskOfPaymentSystems} present the fact that netting brings changes to the network dependence and connectivity, lowering the impact of financial shocks by reducing debts between banks.

Therefore, an efficient netting system generates an important help in reducing the risk of insolvency for companies and, in a broader picture, in reducing the systemic risk and saving the economy from a domino effect~\cite{SYSTEMIC_RISK_IN_INTERNATIONAL_SETTLEMENTS}.%
\paragraph*{\bf Connection with graph theory.}
A netting system can be rendered in terms of graph theory. Every company can be represented as a node in a graph. A directed edge between two nodes designates the fact that one of the companies issued a set of invoices to the other. Given two nodes $u$ and $v$ we may have simultaneously the edges $(u,v)$ and $(v,u)$ , considering that each company has issued invoices to the other. In this situation, the meaning is that a company is not only a customer but also a provider to the counterparty company.
\paragraph*{\bf Motivation.}

The goal of this paper is to present algorithms that improve the netting system in Romania. Our tests use the \emph{real data} provided by the system for debt and credit clearing which runs under the supervision of the Institute of Management and Informatics, the Romanian Ministry of Economy. 

\begin{figure}[!htb]
\centering
\includegraphics[width=1\columnwidth]{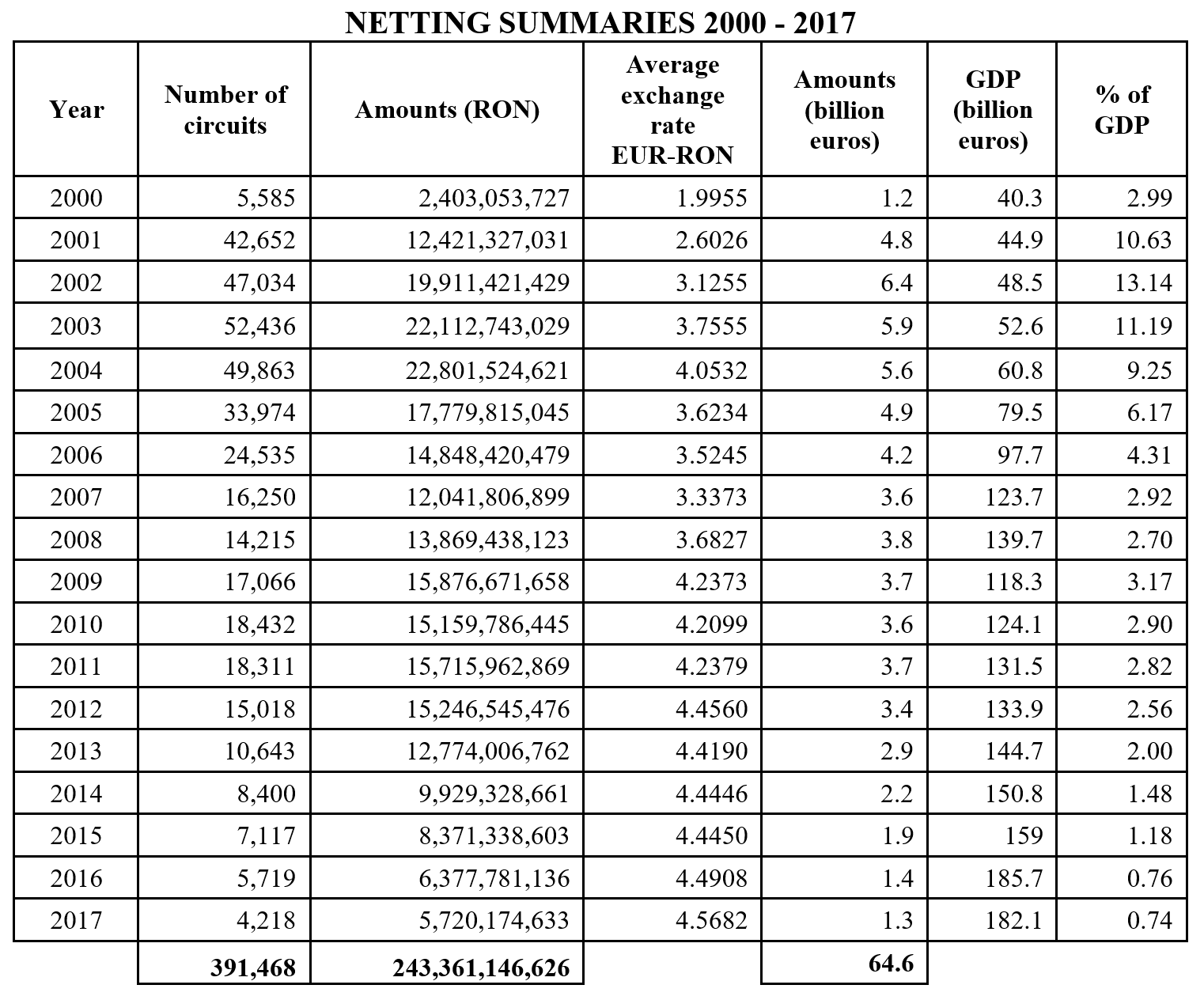}
\caption{Netting amounts  - Romanian Institute of Management and Informatics, 2000 - 2017}
\label{fig:nettingsummaries}
\end{figure}

According to the public data, between 1999 and 2017, a total amount of more than 240 billion RON was netted (more than 60 billion Euros). This amount represents an important percentage of the annual GDP of Romania. One has to reflect on the idea that the cleared amount was based on circuits determined in an empirical manner. The circuits were identified manually, by the fact that the companies knew their partners and proposed clearing solutions. Therefore, the netting solutions were not provided with the aid of an automated system of algorithms. The implementation of an efficient flow of algorithms which run at the level of an economy for an entire country provides great benefits to the economic environment. A table of the public results issued by the Institute of Management and Informatics from Romania is presented in Figure \ref{fig:nettingsummaries}, with more details in~\cite{GamaImiRo}. 

Starting from the 1st of January 2020, the netting platform in the Institute of Management and Informatics (IMI), Romanian Ministry of Economy, was upgraded to use the algorithms proposed in this paper in order to automate the process of circuits computation. In January 2020, more than 100 million Euros were netted by using our computation model. 

\paragraph*{\bf Previous and related work.}
The concept of netting has been described in numerous papers related to financial markets, commerce and economy. Some of the most important results were published in \cite{PaymentsNettingInInternationalCashManagement,OnOptimizingPaymentsClearanceBusinessProcess,EfficientAlgorithmsForTheClearingOfInterbankPayments,ClearingAndSettlementSystemsFromAroundTheWorldAQualitativeAnalysis, SomePracticalPaymentsClearanceAlgorithms}. Netting is mainly involved and generates the best results in interbank payments and in cash optimization for corporation branches in an international environment ~\cite{EfficientAlgorithmsForTheClearingOfInterbankPayments}. G\"untzer,  Jungnickel and Leclerc~\cite{EfficientAlgorithmsForTheClearingOfInterbankPayments} outline the idea that interbank payments represent a process which involves high volumes of participants and large amounts of money, with an example focused on the Germany's interbank payment system.  Srinivasan and  Kim~\cite{PaymentsNettingInInternationalCashManagement} highlight the importance of payment netting between multinational corporations and claim that corporations generate larger and larger volumes of inter-company money flows. Therefore, by implementing an efficient clearing system one can obtain significant savings in payment and interest costs. 
Csoka and Hearings~\cite{Csoka} present the importance of clearing in relation with financial systemic risk and suggest a decentralized clearing method based on a clearing payment matrix. In the literature of financial networks, a set of algorithms for the determination of the greatest clearing payment matrix have been outlined by Eisenberg and Noe~\cite{Eisenberg}, Elliot~\cite{Elliot}, Rogers and Veraart~\cite{Veraart}.

The aforementioned papers motivate the significance of netting in systems like interbank payments or large corporations. As we show in this paper, the concept of netting can be extended to a larger scale. The only application of a clearing system at the scale of a country economy is known to function in Romania. In the Romanian system, the acting parties are represented by every economic agent, regardless of its size.

For solving clearing problems multiple approaches have been proposed. For example,~\cite{EfficientAlgorithmsForTheClearingOfInterbankPayments} presents some simple and efficient heuristic algorithms to accommodate the netting problem. According to~\cite{ShapiroPaymentsNettingInInternationalCashManagement}, Shapiro formulated the multinational payment clearing problem in terms of linear programming, while Srinivasan and Kim present in~\cite{PaymentsNettingInInternationalCashManagement} a network optimization approach considered not only as computationally efficient but also intuitively appealing.  For the netting problem based on inter-company invoices,  Kumlander~\cite{SomePracticalPaymentsClearanceAlgorithms} outlines a set of practical clearance algorithms in based on graph theory. According to~\cite{ClearingAlgorithmsForBarterExchangeMarkets_EnablingNationwideKidneyExchanges}, clearing problems that have bounds on the length of the circuits are NP-hard. Thus, according to the conjecture that $P \neq NP$, polynomial time exact algorithms for this problem are unlikely to exist.

\paragraph*{\bf Our results.}

In this paper, we propose an alternative approach for computing efficient netting solutions. Considering the peculiarities of invoice netting and the fact that the netting circuits length can have an upper limit, we propose a set of graph theory algorithms to find the exact circuits in an invoice graph. It is worth mentioning that our algorithms are already accepted as an upgrade for the Debt and Credit Clearing system running under the Romanian Ministry of Economy authority, the Institute of Management and Informatics. The algorithms are integrated into the netting platform since the 1st of January 2020. In three weeks, the processed sums exceed 100 million Euros. Our paper is the only one proposing a system for debt and credit clearing at the scale of a country economy.

The context of invoice netting has a much larger set of input and output data in comparison with interbank and corporate netting. Thus, an integer programming approach cannot efficiently process such a large amount of data. In our algorithms, the most important part is the circuit identification, based on Johnson's algorithm~\cite{JohnsonsAlgorithm}. According to Leiserson's analysis in~\cite{Leiserson}, Johnson's algorithm is a polynomial time algorithm. It is based on two widely known subroutines, Bellman-Ford (in order to eliminate negative edges) and Dijkstra's shortest path algorithm. The efficiency of our solution, in comparison with heuristic approaches, comes from the fact that the invoice graph is splitted into strongly connected components which reflect the interaction of companies in the economy. By splitting the main problem into smaller similar problems, we obtain an efficient running time. For every connected component, we compute a set of netting circuits. Another advantage of our solution is that, for every set of the netting circuits from a strongly connected component, we compute the best settlement order in terms of maximum clearing amount.


Our algorithm has the following main steps:
\begin{enumerate}
\item represent a directed graph based on invoice information;
\item compute the strongly connected components of the graph using Tarjan's Algorithm;
\item determine the elementary circuits for each strongly connected component by Johnson's Algorithm;
\item compute the netting succession of circuits in order to maximize the total clearing amount.
\end{enumerate}

The first three steps involve algorithms which were covered substantially in research papers (\cite{Tarjan,IntroductionToGraphTheoryWest, JohnsonsAlgorithm,Leiserson}). In this paper, the above mentioned algorithms are adapted for the netting problem. The third step, represented by Johnson's algorithm, is for the first time suggested as a possible computation framework for an invoice netting problem in this paper. The fourth step represents an innovative algorithmic solution for maximization of the total clearing amount. We are aware of the fact that our paper does not use groundbreaking algorithmic techniques. Nevertheless, we find it exciting when modern daily life problems are solved using neat applications of classical textbook algorithms.


The rest of the paper is organized as follows: In Section~\ref{sec:netting} we describe the netting algorithms for our computational model and the implementation details. Then, in Section~\ref{sec:experiments} we present the results of our experiments based on data provided by the Romanian Ministry of Economy. Finally, in Section~\ref{sec:concl} we present conclusions and future research directions.

\section{Netting algorithms}
\label{sec:netting}

In this section we present details related to the proposed netting algorithms. At first, we described the invoice graph representation and the strongly connected components computation. Secondly, we describe the use of Johnson's algorithm for netting circuits computation. Finally, we present the algorithms for netting amounts maximization.

\subsection{Representing a directed graph based on invoice data and computing the strongly connected components} \label{ssec:num2.1}

A directed graph is an ordered pair $G = ( V, E )$ (see~\cite{IntroductionToGraphTheoryWest} for basic graph terminology). The set $V$ represents a set of vertices also known as nodes or points. The set $E$ is defined as a set of ordered pairs of nodes, known also as directed edges, arcs or directed lines. 
Considering the netting model, the companies are represented as nodes in the graph. A payment obligation from a company to another company is rendered as an edge in the graph. For example, if company \textit{A} issues a set of invoices to company \textit{B}, an edge from \textit{B} to \textit{A} has the total weight of the sum of invoices. 
 One simple circuit (i.e., no vertices nor edges can repeat) in the graph represents a set of companies for which a netting can be implemented. 

\begin{definition}[Netting Problem]
The input of the problem is a directed graph G = (V, E), with a cost $w : E \to \mathbb{R}$ associated to each edge. The operation of \emph{settling} a circuit consists of decreasing the value of all the edges in the circuit with the minimum value of an edge in that circuit. That is, given a circuit $(v_1, v_2, \dots v_k, v_1)$, let $x$ be the minimum value of an edge in the circuit, i.e., $min ( $$min_{i=1}^{k-1} w(v_i,v_{i+1}), w(v_k, v_1) )$ by replacing the weight for each edge $w(v_i,v_{i+1})$ with $w'(v_i,v_{i+1}) = w(v_i,v_{i+1}) - x$, respectively $w'(v_k,v_1) = w(v_k,v_1) - x$.

The goal of the netting problem is to find an ordered set of circuits such that settling them in the prescribed order maximizes the total netted amount.

\end{definition}

According to~\cite{ClearingAlgorithmsForBarterExchangeMarkets_EnablingNationwideKidneyExchanges}, the netting problem defined above is NP-hard. Johnson's algorithm, as it is described in \cite{JohnsonsAlgorithm}, represents a method of computation for the all pairs shortest path problem considering a directed sparse graph. The graph density $D$ is defined as $|D| = \frac{|E|}{|V|(|V| - 1)}$.

Recall that, in our case, the nodes in the invoice graph are represented by companies while the edges represent invoice relations between companies. In an economy, a company interacts with a limited number of counterparties and this gives us the intuition of why the invoice graph is a sparse one. Considering the graph density formula, we made simulations on the invoice graphs created from real data provided by Romanian Institute of Management and Informatics and, indeed, we observe that the invoice graph is a sparse one. 

\begin{figure}[htb]
\centering
\includegraphics[width=\columnwidth]{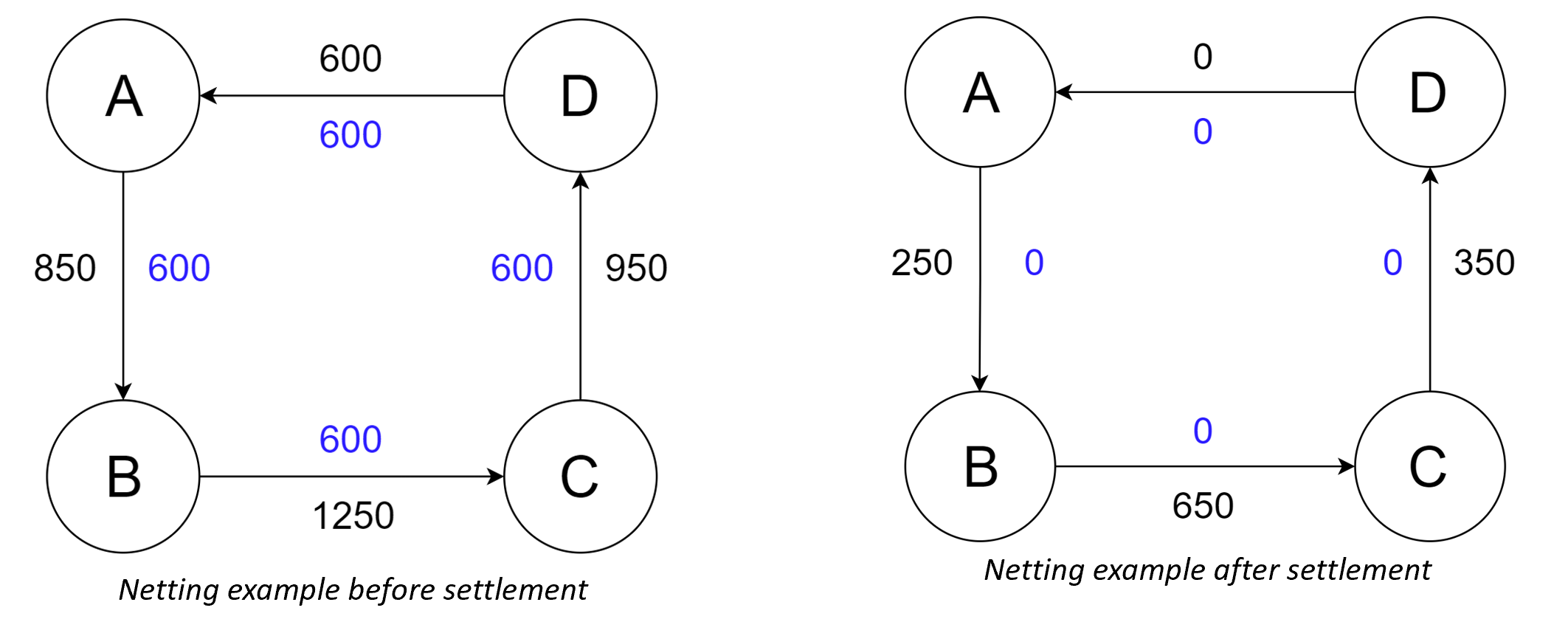}
\caption{\label{fig:NettingImplementationExample}Netting implementation example
}
\end{figure}

Companies upload their invoices in the netting system. At a regular interval the graph is updated, the circuits are computed and the companies are notified about possible nettings. If all the companies in a netting circuit approve the settlement, the sums are discarded from the graph. In this manner, the companies diminish their mutual obligations. As an example, Figure~\ref{fig:NettingImplementationExample} outlines a netting scenario before and after netting implementation. Considering four companies (A, B, C, D) as vertices, the liabilities between them are marked on the edges. The minimum value of an edge in the ABCD circuit is 600 (the edge between D and A). Therefore, a netting of 600 can be implemented between all of these companies. After the netting implementation,  600 is subtracted from the value of every edge. The liability of D for A becomes 0. The liabilities marked on the edges AB, BC, CD have positive value, and can take part in other netting circuits.

The first step in computing the netting circuits consists in identifying the strongly connected components of the graph of invoices. The approach in this paper is to consider Tarjan's algorithm, which was firstly introduced in~\cite{Tarjan}. A directed graph is strongly connected if there is a path between all pairs of vertices (see~\cite{IntroductionToGraphTheoryWest} for a formal definition). A strongly connected component (SCC) of a directed graph is a maximal strongly connected subgraph. Because the invoice graph is a sparse one, in order to efficiently determine the circuits, we have to find  first the strongly connected components of the graph, and start the circuit computation on each of these  components. In order to determine the SCC, a choice had to be made between the Kosaraju's algorithm and Tarjan's algorithm. As it is detailed in~\cite{Tarjan}, considering the fact that Kosaraju's algorithm requires two DFS traversals of the graph, while Tarjan's algorithm involves only one DFS traversal, the latter one was considered. The time complexity for Tarjan's algorithm is $\mathcal{O}(|V| + |E|)$. Tarjan algorithm is based on following ideas (see Algorithm~\ref{alg:tarjan} for a detailed description):
\begin{itemize}
\item a DFS tree/forest is obtained through a DFS search;
\item SCCs represent subtrees of the DFS tree;
\item starting from the head of a subtree, all the other nodes in the SCC can be obtained;
\item there is no back edge from one SCC to another. (There can be cross edges, but cross edges will not be used while processing the graph).
\end{itemize}

\begin{algorithm}[htb] \caption{Tarjan's algorithm for determining the strongly connected components}
\label{alg:tarjan}
\large
\begin{algorithmic}[1]
\Function{StrongConnect}{vertex $V$}
\State $num \leftarrow num + 1$ \;
\State $order[V] \leftarrow num$ \;
\State $link[V] \leftarrow order[V] $ \;
\State push $V$ on the $Stack$ \;

\For{ $succesor$ in $vertexSuccesorsOfV$ } {
  \If{$order[succesor]$ is not defined}
    \State StrongConnect($succesor$) \;
    \State $link[succesor] \leftarrow min(link[V], link[succesor])$ \;
  \Else \If{$V$ is on the $Stack$} 
          \State $link[V] \leftarrow min(link[V], order[succesor])$ \;
        \EndIf
  \EndIf
  }
\EndFor

\If{$link[currentV] = order[currentV]$}
  \State Create a new strongly connected component \;
  \Repeat
  	\State $auxVertex \leftarrow top Of Stack$
    \State add $auxVertex$ to strongly connected component
    \State pop top from the $Stack$
  \Until {$currentV = auxVertex$}
 \EndIf
\EndFunction

\Function{TARJAN}{graph $G$}
	\State $num \leftarrow 0$ \;
    \State initialize $Stack$ \;
    \For {vertex $V$ in graph $G$} {
    	\If {$order[V]$ is undefined}
        	\State StrongConnect($V$)
        \EndIf
    }
    \EndFor
\EndFunction
\end{algorithmic}
\end{algorithm}

\subsection{Johnson's algorithm for netting circuits computation} \label{ssec:num2.2}

Johnson's algorithm, first published in~\cite{JohnsonsAlgorithm}, is a method of computation of all pairs shortest path problem in a directed sparse graph. Our problem requires a particular approach for the Johnson's algorithm, considering the fact that all the weights are nonnegative. For the circuit enumeration problem in directed sparse graphs, the solution proposed by Johnson consists in a polynomial time algorithm. As Leiserson analyzed in \cite{Leiserson}, Johnson's algorithm is very similar to the Floyd-Warshall algorithm. In terms of efficiency, Johnson's algorithm time complexity is in direct dependency with the number of arcs in a graph, as opposed to Floyd-Warshall, which has a different approach (direct dependency with the number of nodes in a graph). This is the reason why the algorithm developed by Donald B. Johnson is more suitable for sparse graphs, while Floyd-Warshall is more effective on dense graphs. Johnson's algorithm runs in $\mathcal{O}(V^2\log{}V + |V||E|)$ time, while Floyd-Warshal in $\mathcal{O}(V^3)$. As Leiserson outlines in \cite{Leiserson}, Johnson's algorithms is based on two well known shortest path subroutines. The first one is Bellman-Ford, used in order to eliminate negative edges and identify negative circuits and to reweight the initial graph. The second one is Dijkstra's shortest path algorithm used to determine the shortest path between all pairs of nodes. In our case, considering the fact that the value of edges can only be positive, reflecting the sum of invoices, the Bellman-Ford step is not necessary.

\begin{algorithm}[h!tb] \caption{Variation of Johnson's algorithm for computing the netting circuits}
\large
\begin{algorithmic}[1]
\Function{CIRCUIT}{vertex $V$}

\State $circuitFound\leftarrow false$    \;
\State $maxNumberOfCircuitComponentsReached\leftarrow false$    \;
\State push $V$ in the $stack$ \;
\State $blockedVrtxList[V]\leftarrow true$ \;

\If{$sizeOf(stack) \geq maxNumberOfCircuitComponents$}
  \State $maxNumberOfComponentsReached\leftarrow true$    \;
\Else
\For{ $vertexSuccesor$ in $succesorsOfV$ } {
\If{$vertexSuccesor$ is $startVertex$}
  \State Save current circuit \;
  \State $circuitFound\leftarrow true$    \;
\Else
  \If {$blockedVrtxList[vertexSuccesor]$ is false}
  	\State $circuitFound \leftarrow CIRCUIT(vertexSuccesor)$ \;
  \EndIf
\EndIf
}
\EndFor
\EndIf

\If{$circuitFound$ or $maxNumberOfCircuitComponentsReached$}
	\State UNBLOCK($V$);
\Else
	\For{ $vertexSuccesor$ in $succesorsOfCurrentVertex$ }
    {
    	\If{$V$ is not in $blockedListForVertex$[$vertexSuccesor$]}
        	\State Add $V$ to $blockedListForVertex$[$vertexSuccesor$] \;
        \EndIf
    }
    \EndFor
\EndIf
	
\State pop from the $stack$

\State return $circuitFound$

\EndFunction
\end{algorithmic}
\end{algorithm}

\begin{algorithm}[htb] \caption{Unblock procedure}
\large
\begin{algorithmic}[1]
\Function{Unblock}{vertex $V$}

\State $blocked[V]\leftarrow true$ \;
\While {$blockedListForVertex[V]$ is not empty} 
	\State $vertexToUnblock \leftarrow$ first element from $blockedListForVertex[V]$
    \If {$vertexToUnblock$ is in $blockedVertexesList$}
    	\State UNBLOCK($vertexToUnblock$) \;
    \EndIf
\EndWhile
            
\EndFunction
\end{algorithmic}
\end{algorithm}
In our variation of Johnson's algorithm, at line 6 one can notice the use of $maxNumberOfCircuitComponents$. This is the adaption for the netting problem we made from the original Johnson's algorithm. The circuit search is stopped if the stack exceeds this threshold value. The idea is that a netting circuit has to be approved by all of the members, in order to be effectively implemented. The longer a circuit, the lower the probability of acceptance for all the  members. Therefore, the search is limited to a certain variable. In order to set the value of the $maxNumberOfCircuitComponents$, a statistic is realized for computing the most probable size of an implemented circuit. Therefore, the variable is set as twice the value of the most probable circuit length.

\subsection{Maximize the netting implementation}

\begin{figure}[h!tb]
\centering
\includegraphics[width=\columnwidth]{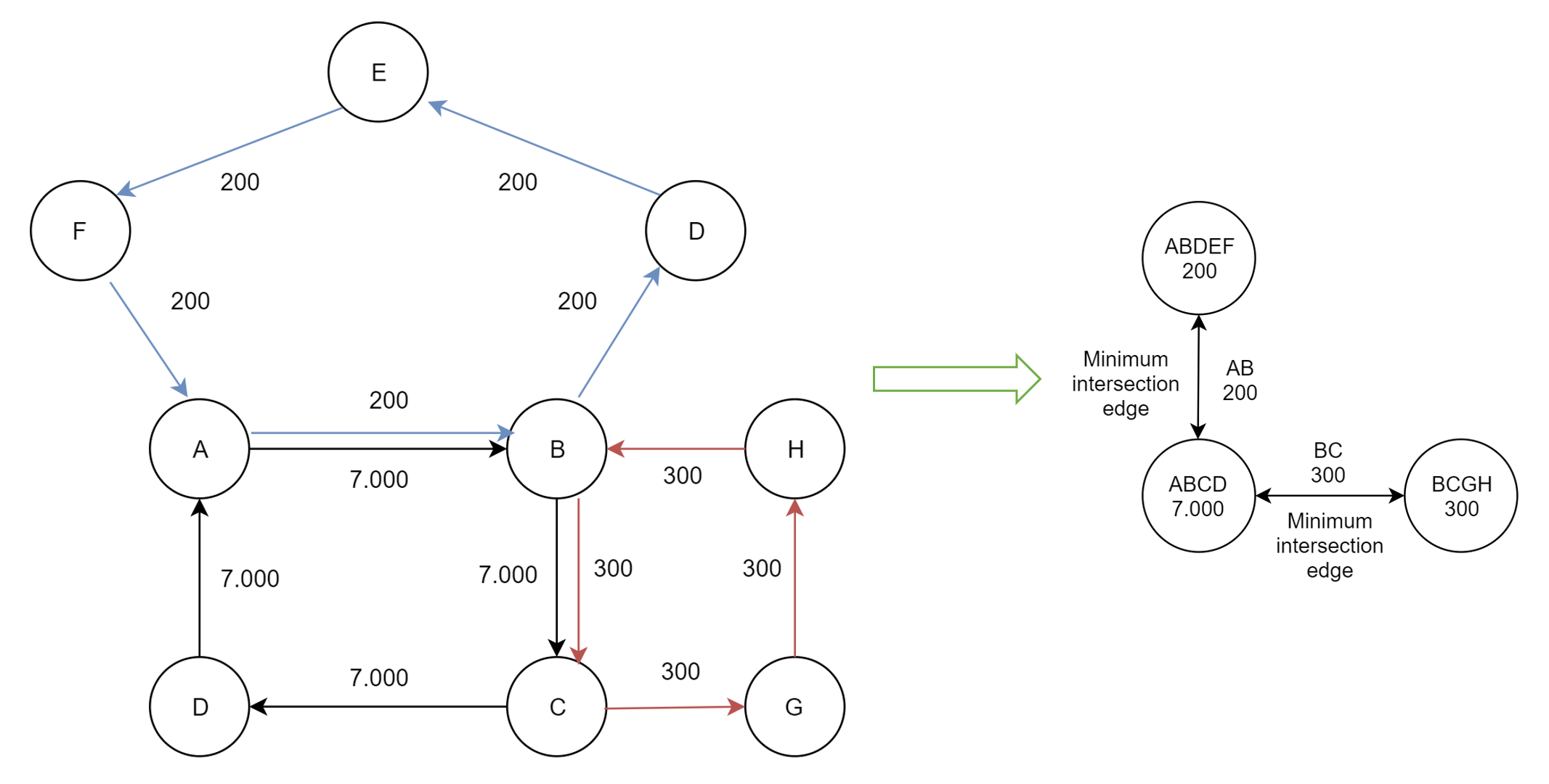}
\caption{\label{fig:CircuitSettlementOptimizationExample}Circuit settlement optimization example}

\end{figure}

The algorithms outlined in Subsection \ref{ssec:num2.1} and Subsection \ref{ssec:num2.2} present how the netting circuits are computed. Once we have these circuits, another important problem is to maximize the total netting implementation value. The maximization can be realized by selecting a proper order for netting. For example, Figure \ref{fig:CircuitSettlementOptimizationExample} outlines three netting circuits. If one firstly implements circuit ABCD, a total amount of 28.000 ($7000 \times 4$) is obtained. In this situation, the second circuit ABDEF cannot be implemented, because the edge AB would have a remaining value of 0. Moreover, the circuit BCGH cannot be implemented, because the edge BC would have a remaining value of 0. However, if the ABDEF circuit is implemented  first, an amount of 1000 is generated ($200 \times 5$). Thereafter, for the ABCD circuit an amount of $27\,200$ is generated. Considering both circuits, ABDEF and ABCD, in this order or  implementation, a total amount of $28\,200$ is generated. In this case, circuit BCGH cannot be implemented. Similarly, the edge BC would have a remaining value of 0. The optimal implementation order is ABDEF, BCGH and ABCD. This implementation results in a total amount of $29\,000$. 

The aforementioned example is a simple one, considering just three circuits. In real life, hundreds of thousands circuits have to placed in an optimal implementation order. Even if in the previous example the total gain seems to be a small one, in other situations the gain is significantly higher. We describe an algorithm for circuit implementation optimization. Every circuit computed in the previous section represents a vertex in a graph. In this graph, an edge between two vertices represents the fact that the two circuits have an intersection. The value of the edge is considered as the smallest value of a segment in the intersection.

Therefore, if one considers vertex 1 and vertex 2, if the edge between them has value 200, it means that the smallest intersection between the two circuits is 200. Therefore, if a netting is implemented by using circuit 1, and if the netting value per segment is 200, then the second circuit cannot be implemented, because the segment of 200 is already consumed. However, if the netting is implemented by using circuit 1, and if the netting value per segment is 150, therefore the intersection between circuit 1 and circuit 2 is reduced to 50. Therefore, considering circuit 2, the maximum amount which can be considered in netting per segment is the minimum between 50 and the minimum segment of circuit 2 outside of the intersection with circuit 1. This idea is outlined by Figure~\ref{fig:CircuitSettlementOptimizationExample}, in which we have as nodes the previously presented circuits, ABCD (with an amount of 7000 per edge), ABDEF (with an amount of 200 per edge), BCGH (with an amount of 300 per edge). 
As we previously mentioned, the flow of algorithms is: create the invoice graph; compute the strongly connected components; compute the circuits for every strongly connected components; optimize the netting implementation order. In this optimization phase, the algorithm runs on every set of circuits resulted from a strongly connected component. More precisely, for every set of circuits resulted from a strongly connected component, we select an arbitrary vertex $v$ and call \textsc{OptimizationAlgorithm} on $v$ (see Algorithm~\ref{alg:circuit order}). Algorithm~\ref{alg:circuit order} exhaustively tries all the possible ordering of the circuits, which guarantees its optimality.


\begin{algorithm}[h!tb] \caption{Circuit settlement optimization}
\label{alg:circuit order}
\large
\begin{algorithmic}[1]
\Function{OptimizationAlgorithm}{vertex $V$}
\State Add V to settlement solution
\State Eliminate from the graph all vertices which	are excluded because of the implementation of the $V$

\If {there are any nodes remaining in Graph} {
\For {vertex $next$ in Graph}
  \State OptimizationAlgorithm($next$)
\EndFor
}
\Else
  \State Save the optimization solution and its value \;	
\EndIf
\State Remove current node from the settlement solution \;
\State Reload into the graph the nodes excluded previously because of the implementation of current node \;
            
\EndFunction
\end{algorithmic}
\end{algorithm}

\section{Experiments}
\label{sec:experiments}

In this section we outline details regarding our experiments. Also, we present several interesting implementation details. Our flow of algorithms has four phases: directed graph creation, strongly connected components computation, circuits determination, order of circuits implementation optimization. The flow of algorithms was tested on raw data provided by the Institute of Management and Informatics, Romanian Ministry of Economy. The algorithms were implemented by using C\# .NET Core 3.0, a Microsoft Technology. The hardware involved in the experiments consisted in two Dell PowerEdge R730 Servers, with 128 GB of RAM. 

We tested our algorithm on both real data provided by the Romanian Ministry of Economy in order to compare our algorithms with the results achieved manually the period 2000-2017. Moreover, we also use some simulated data which contains much larger instances than the real ones. Thus, we aim to test the robustness of our algorithm.





\begin{figure}[h!tb]
\centering
\includegraphics[width=0.6\columnwidth]{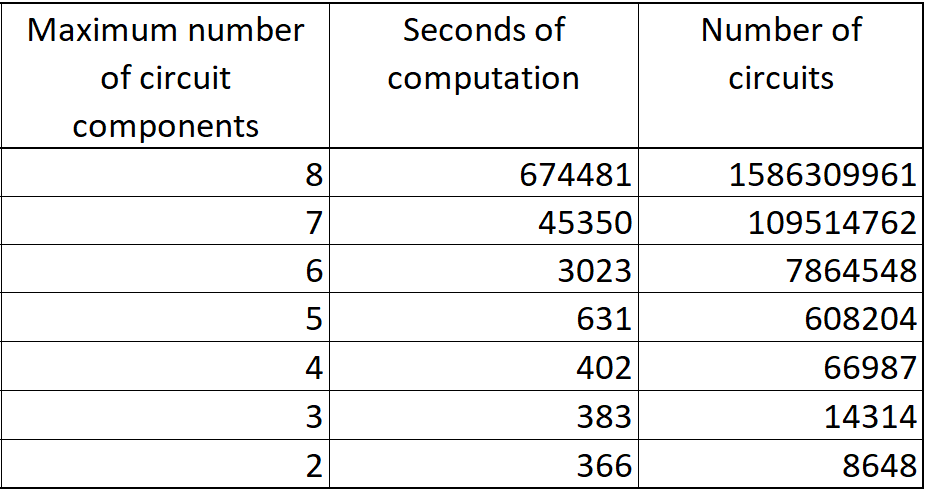}
\caption{\label{fig:GrafPentru50000}Experiment results. }
\end{figure}

Then, in Figure~\ref{fig:GrafPentru50000}, we outline an example of computation in which $15\,000$ companies are involved, with $50\,000$ distinct invoice paths between them. We carried out several experiments, considering the computation of circuits between 2 and 8 components. From the statistics provided by the Institute of Management and Informatics, more than 90\% of the nettings are realised by circuits with maximum 4 components. In our simulations we have a circuit length of maximum 8 components, double the size of the most probable circuit length, which is 4. The running time for computing circuits of maximum 8 components is a feasible one for the practical application. As we observe from the test data set of $15\,000$ companies and $50\,000$ distinct invoice paths,  computing circuits with more than 6 components conducts in a considerable increase of computation time.

\begin{figure}[h!tb]
\centering
\includegraphics[width=1\columnwidth]{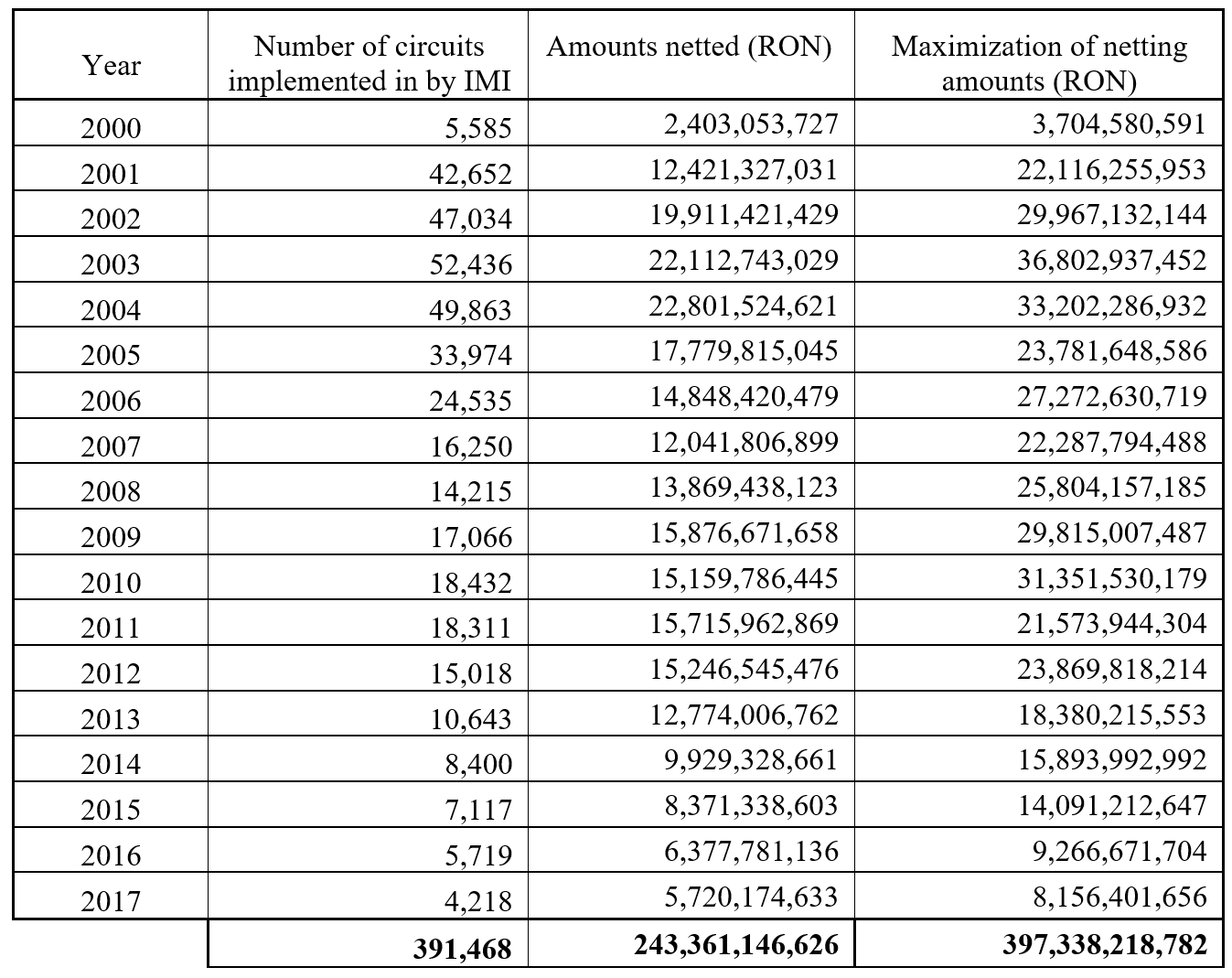}
\caption{\label{fig:MaximizationOfNettingAmunts}Comparison between the amounts netted by IMI in  comparison with the results of applying the maximization algorithm}
\end{figure}
\begin{figure}[h!tb]
\centering
\includegraphics[width=1\columnwidth]{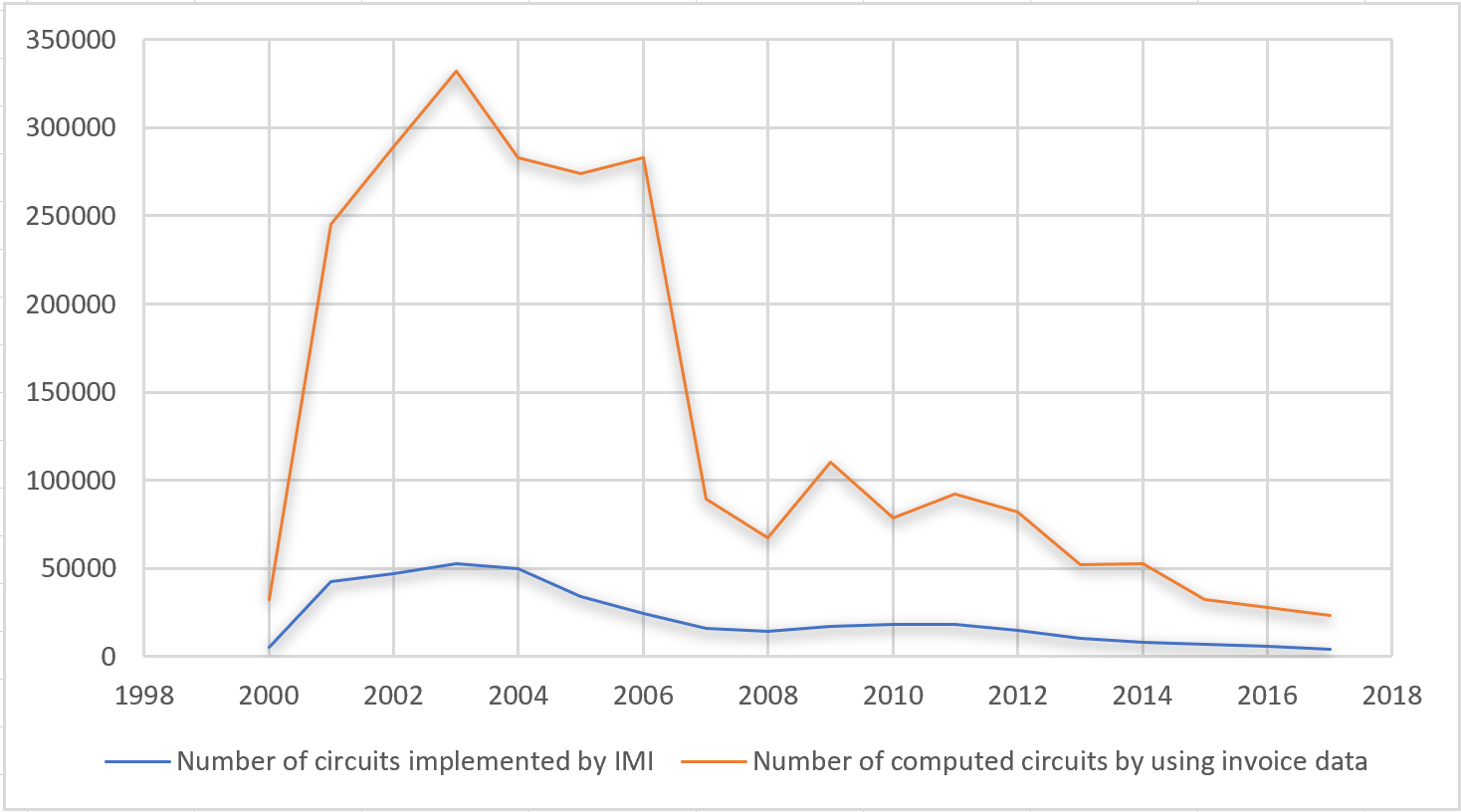}
\caption{\label{fig:GraphForNumberOfComputedCircuitsByUsingInvoiceData}Graph for the number of circuits implemented by IMI, compared with the number of circuits computed by the flow of algorithms for the invoices implied by the implemented circuits}
\end{figure}

We compare in Figures~\ref{fig:MaximizationOfNettingAmunts} and~\ref{fig:GraphForNumberOfComputedCircuitsByUsingInvoiceData} the number of netting circuits implemented by IMI, Ministry of Economy with the number of circuits computed by the flow of algorithms (first three phases: directed graph creation, strongly connected components computation, circuits determination) for the invoices used in the implemented nettings. In order to compute the number of circuits for every year, we consider the invoices on which the real life nettings were realised in that year. For example, in 2004, IMI implemented $49\,863$ nettings. Based on the invoices which support these nettings our algorithms determine $282\,727$ circuits. Notice that the total number of computed circuits is at least five times the number of implemented nettings. Recall the fact that IMI implemented circuits as a request from the business agents, based on their day by day interaction, circuits not resulted from a method of computation. Therefore, these figures outline the fact that by using the flow of algorithms the set of netting solutions would increase significantly. Moreover, by registering more and more invoices, the set of computed netting circuits would increase considerably.

\begin{figure}[!htb]
\centering
\includegraphics[width=\columnwidth]{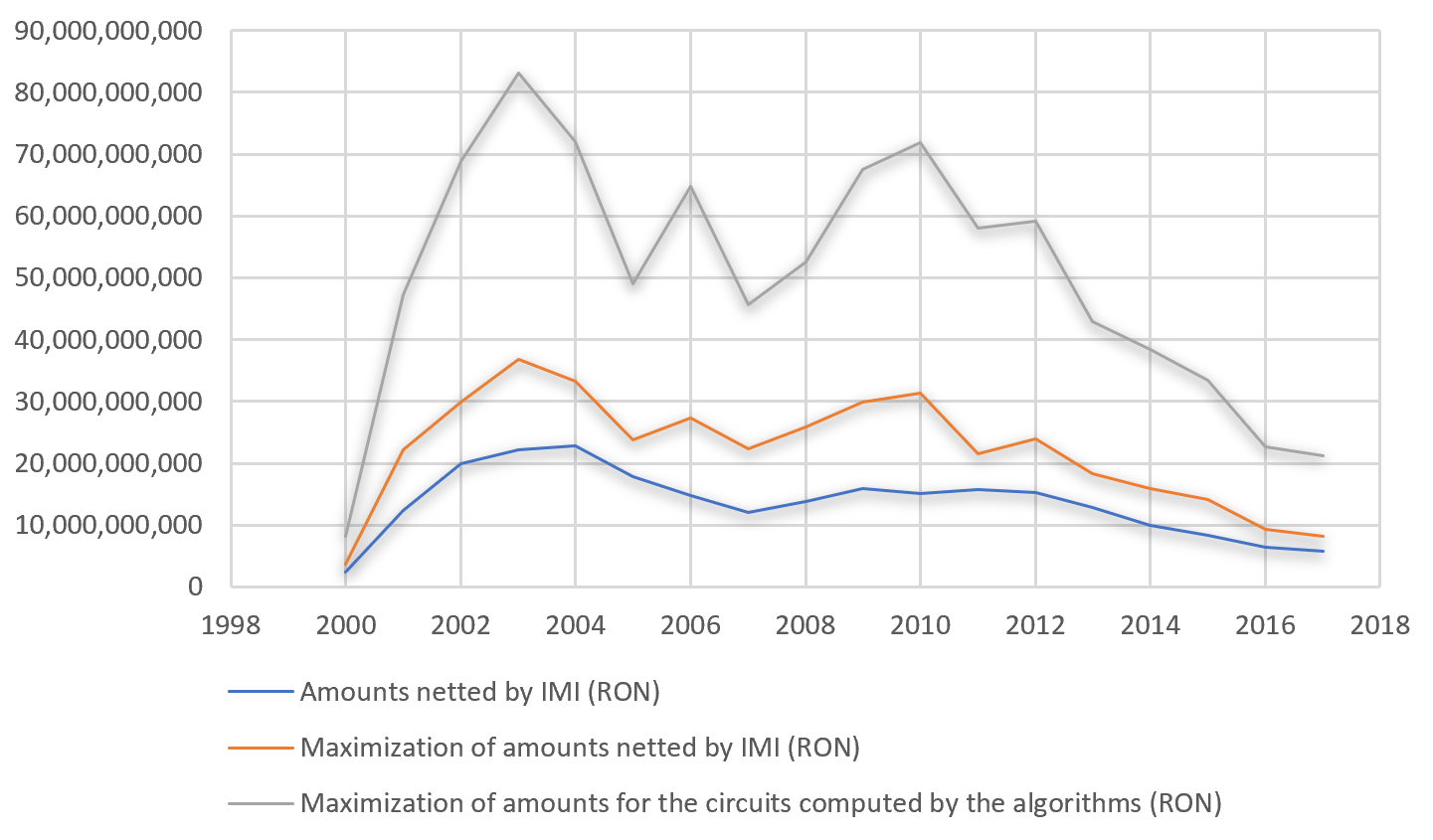}
\caption{\label{fig:GraphMaximizationOfNettingAmounts}Graph for comparing the netting amounts realised by IMI, the results of the maximization algorithm phase for the netting circuits implemented by IMI, the results of the maximization  algorithm phase for the netting computed circuits}
\end{figure}

In Figures~\ref{fig:MaximizationOfNettingAmunts} and~\ref{fig:GraphMaximizationOfNettingAmounts} we outline the results of applying the fourth step of the algorithm flow (order of circuits implementation optimization) for the netting circuits implemented by IMI, but also for the sets of computed circuits. One can notice the fact that the sum of netted amount would have increased from around 243 billion to around 397 billion if the proper order of implementation would have been used for the nettings realised by IMI (the increase is more than 60\%). Moreover, if the computed circuits would have been implemented in the proper order, a total amount of around 907 billion RON would have been realised. The sum represents an increase of around 270\% in comparison with the amount of 243 billion implemented by IMI.




\section{Conclusion}
\label{sec:concl}

Clearing represents an important mechanism for improving the  interactions between the companies in an economy. It consists in an act of agreement in which each party sets off amounts it owes against the amounts owed to it. In this paper we presented a set of graph algorithms in order to compute netting solutions at the economy scale of a country. The analysis and experiments are realized based on real life data provided by the Romanian Ministry of Economy.

As future work, we aim to consider an additional type of optimization in order to maximize the total number of participants in the netting process. In the current paper we optimized the maximum netted amount. However, another helpful aspect for the economy is to involve as many companies as possible in order to benefit from the advantage offered by clearing. By maximizing the netting amount, we noticed that in some cases smaller and medium sized companies which interact with large ones are disadvantaged, because the proposed circuits tend to advantage the larger companies in order to maximize the sums. We plan design an optimization algorithm for balancing the netted amount and the number of participants.


\bibliographystyle{abbrv}
\bibliography{bibliography}

\end{document}